\documentclass[11pt]{article}
\usepackage{moriond,epsfig}

\usepackage{graphicx}

\def\be{\begin{equation}}
\def\ee{\end{equation}}
\def\bea{\begin{eqnarray}}
\def\eea{\end{eqnarray}}

\newcommand{\ba}{\begin{eqnarray}}
\newcommand{\ea}{\end{eqnarray}}

\begin{document}
\vspace*{4cm}	
\title{OPTICAL CLOCK AND DRAG-FREE REQUIREMENTS\\
 FOR A SHAPIRO TIME-DELAY MISSION}

\author{N. ASHBY$^1$ \& P. BENDER$^2$}

\address{$^1$ University of Colorado, Boulder, CO \\ 
$^2$JILA, University of Colorado, Boulder, CO}

\maketitle\abstracts{
  In the next decade or two, extremely accurate tests of
general relativity under extreme conditions are expected from gravitational
wave observations of binary black hole mergers with a wide range of mass
ratios.  In addition, major improvements are planned in both strong and weak equivalence principle tests; clock measurements based on
the ACES program on the ISS; more accurate light-bending measurements; and
other new types of tests.  However, whether these tests are all consistent
with general relativity or not, it still appears desirable to proceed with a
much improved measurement of the Shapiro time delay.
  A suggested approach\cite{ashby10} is based on
using a high-quality optical clock in a drag-free spacecraft near the
sun-earth L1 point and a smaller drag-free transponder spacecraft in a two-year period
solar orbit.  Laser phase travel-time measurements would be made between the
two spacecraft over a period of 10 or 20 days around the time when the line
of sight passes through the Sun.  The requirements on the optical clock
stability and on the drag-free systems will be discussed.  The accuracy achievable for the time-delay appears to be
better than 1 part in 100 million.}

\section{Introduction}

The first suggestion to measure the gravitational time delay for electromagnetic waves passing near the sun was made by Irwin I. Shapiro in 1964.\cite{shapiro64}  The extra gravitational time delay for two-way measurements of light propagating from Earth to a spacecraft passing behind the Sun can be more than 200 microseconds.  In the Parametrized Post-Newtonian (PPN) formulation of gravitational theory, the main contribution to the time delay is proportional to $(1+\gamma)$, where $\gamma$ is a measure of the curvature of space. In General Relativity (GR), $\gamma=1$.

In view of the well-known lack of a theory that connects GR with quantum theory, improvement of high-accuracy tests of the predictions of GR should be the object of research in the coming decade.  Many alternatives to GR involve additional scalar fields.  Studies of the evolution of scalar fields in the matter-dominated era of the universe indicate that the universe's expansion tends to drive the scalar fields toward a state in which the scalar-tensor theory is only slightly different from GR.  Some scalar-tensor extensions of GR\,\cite{damour93,damour96} predict deviations from the GR value of $\gamma$ in the range from $10^{-5}$ to $10^{-8}$.  Improved information about $\gamma$ would provide important insight into the evolution of the universe and directly limit the range of applicabilituy of alternative gravitational theories.

Recently, a measurement of $\gamma$ with accuracy $\pm 2.3 \times 10^{-5}$ was made during the Cassini mission.\cite{bertottietal03}  Further improvements in the accuracy for $\gamma$ to roughly $10^{-6}$ are expected from two missions of the European Space Agency (ESA):  the GAIA astrometric mission, which will measure the gravitational deflection of light rays by the sun, and the Bepi Colombo mission to Mercury, which will make improved measurements of the solar time delay.  We describe here a mission\cite{ashby10,bender_etal08,Ashby_Bender08}
 that can reach an accuracy of about $1 \times 10^{-8}$ for determining $\gamma$.

\section{Mission orbits and predicted time delay}

For the proposed mission, one spacecraft (S1) containing a highly stable optical clock would be placed in an orbit near the L1 point, about 1.5 million km from the Earth in the direction of the Sun.  The second spacecraft (S2) would have a 2 year period orbit in the ecliptic plane, with an eccentricity of 0.37.  S2 would pass through superior solar conjunction about 1, 3, and 5 years after launch and would be near aphelion at those times.  Both spacecraft would have drag-free systems to nearly eliminate the effects of spurious non-gravitational forces.  A measurement of $\gamma$  to a level of $1 \times 10^{-8}$  would be carried out by observing the time delay of laser signals exchanged between the two spacecraft when the line of sight passes near the Sun's limb.  Atmospheric effects would be absent and continuous  observation would be possible.  With S2 near aphelion, the range rate would be low, and the orbit determination problem would be much reduced.  

	The crucial measurements of time delay occur within a few days of superior conjunction and are primarily characterized by a logarithmic dependence on the distance of closest approach of the light to the mass source.  The predicted gravitational time delay due to a non-rotating mass source, expressed in terms of the radii $r_A,r_B$ of the endpoints of the photon path, and the elongation angle $\Phi$ between the radius vectors from the source to the endpoints, is \cite{ashby_bertotti} 
\be\label{timedelay}
c\Delta t_{delay}=\mu(1+\gamma)\log\bigg(\frac{r_A+r_B+r_{AB}}{r_A+r_B-r_{AB}} \bigg)-\frac{\mu^2(1+\gamma)^2r_{AB}}{r_A r_B (1+\cos\Phi)}
+\frac{\mu^2 r_{AB}(8-4\beta+8\gamma +3 \epsilon)}{4 r_A r_B \sin \Phi}
\ee
where $\mu=GM_{\odot}/c^2$, and $r_{AB}$ is the geometric distance between the endpoints in isotropic coordinates and $\beta$ and $\epsilon$  are PPN parameters measuring the nonlinearity of the time-time and space-space components of the metric tensor. In GR, $(8-4\beta+8\gamma +3 \epsilon)/4=15/4 $. The time delay in Eq. (\ref{timedelay}) is expressed in terms of observable quantities, and does not involve the unknown impact parameter or distance of closest approach.  The non-linear terms are a few nanoseconds so they are significant, but do not have to be estimated with great accuracy.  The contributions to time delay due to the solar quadrupole moment are small and can be estimated with sufficient accuracy that they will not contribute significantly to the error budget. 

	The measurements will be made by transmitting a laser beam with roughly 40 GHz sidebands on it from from S1 to S2, and comparing with a similar beam generated on S2 and sent back to S1.  From the phase differences of the sideband beat notes, the round-trip delay time can be obtained. 
With 20 cm diameter telescopes, and given the one-way travel time of about 1600 s, the received signal would be roughly 1000 counts/s for 1 W of transmitted power.  This is a weak signal, but it is strong enough so that the chances of a cycle slip should be very small.  If we consider the round-trip delay times $\Delta t_{delay}$ to be the observable, then the change in delay from 0.75 days to 4 days on either side of conjunction is about 64 microseconds.

\section{Signal-to-Noise Analysis}  We can estimate the lowest possible uncertainty that could be attained in this experiment on the basis of the optimal Wiener filter, which takes advantage of the known time signature of the signal and includes the expected clock noise.\cite{ashby10,Ashby_Bender08,thorne87}  For this case, uncertainties in the various orbit parameters for the two spacecraft are ignored, and the travel time between the spacecraft is assumed to be constant except for changes in the gravitational time delay.  The time signature of $\gamma^*=(1+\gamma)/2$ is taken to be represented by the logarithmic function
\be
g(t)=-B (\log\vert R t \vert -M)
\ee
where $M$ is the mean value of $\log\vert Rt\vert$  over the time periods $-t_2$  to $-t_1$  and $t_1$
 to $t_2$ (a short time interval during occultation is excluded), and for the proposed experiment $B=0.97\times 8 \mu /c =3.82\times 10^{-5}$ s.  The rate at which the line of sight to the distant spacecraft passes across the sun is $R = 1.9$ solar radii per day. 

	Let $g(f)$ be the Fourier transform of $g(t)$ over the time of the measurements.  Then the signal-to-noise ratio may be found \cite{ashby10,Ashby_Bender08,thorne87} in terms of an integral over all frequencies of $\vert g(f) \vert^2.$  An important consequence of the logarithmic form of the time delay, Eq. (2), is that if the noise has a constant spectral density, only about 2.5\% of the signal-to-noise ratio comes from frequencies below 1 microHz, where the acceleration and clock noise are expected to increase.  Just integrating down to 1 microHz, we find an uncertainty less than  $1 \times 10^{-9}$ for $\gamma$. Almost all of the power in $\vert g(f) \vert^2$ is at frequencies between 1 and 8 microHz, so it is clear that the noise at these low frequencies will provide the main limitation on the results.

	Actually, all of the in-plane parameters for the orbits of the two spacecraft have to be solved for, as well as $\gamma$.  Our model for this includes uncorrelated 0.02 picosecond uncertainties for measurements of the round-trip travel time over 3-hr periods.  This is in addition to our assumed white clock frequency noise of $5 \times 10^{-15}/\sqrt{{\rm Hz}}$ down to at least 1 microHz.  Spurious acceleration noise is not included in this model, but its effect has been estimated to be small with our assumptions about its spectrum.  The resulting uncertainty in $\gamma$ is less than $1 \times 10^{-8}$. 

	So far, we have assumed that time-delay measurements are only made over a total period of 8 days around solar conjunction.  This was done in order to make sure that spurious acceleration noise at frequencies below 1 microHz would have little effect.  However, simulations for longer observing times are desirable, with full allowance for spurious acceleration noise at the
lowest frequencies, as well as for the orbit determination part of the
problem.  The longer observation period may help to improve the determination of orbit parameters as well as $\gamma$.

\section{Spacecraft S1 clock}

The major requirement for the mission is to fly an optical clock on S1 that
has very high stability over a period of at least 8 days around superior
conjunction.  The nominal design goal for the mission is to achieve a
fractional frequency noise power spectral density amplitude of $5 \times 10^{-15}/\sqrt{{\rm Hz}}$
from 1 Hz down to at least 1 microHz.  (This is nearly equivalent to an Allan
deviation of $5 \times 10^{-15}/\sqrt{\tau}$ for times from 1 s up to $10^6$ s.)

As an example of the desired performance, a spectral amplitude of about
$2 \times 10^{-15}/\sqrt{{\rm Hz}}$ has been achieved in the laboratory down to 1 mHz for the
267 nm transition in sympathetically-cooled Al$^+$ ions in a magnetic trap.\cite{chou10}
Other leading candidates for optical clocks in space are cooled Sr$^{88}$
atoms\cite{swallows11,ludlow08} and Yb$^{171}$ atoms\cite{lemke09,jiang11} in optical latices.  However, substantial
development is needed to show that such optical clocks can be designed for
use in space and can be space qualified.

\section{Drag-free system}

  The required performance builds on that planned for the LISA mission.  For frequencies down to $10^{-4}$ Hz for LISA, the requirement on the acceleration power spectral density amplitude is less than $3\times 10^{-15} {\rm m/s}^2/\sqrt{{\rm Hz}}$.  However, the performance is expected to degrade at lower frequencies.  The main challenge for achieving good performance at low frequencies is minimizing thermal changes, and particularly thermal gradient changes, near the freely floating test mass in the drag-free system. On LISA this is done almost completely by passive thermal isolation.  For a time delay mission, a fairly slow active temperature control system would be used at frequencies below  $10^{-4}$ Hz.  Changes in solar heat input over the 8 days around conjunction would be quite small for S2, because conjunction occurs near aphelion.  The required drag-free performance is roughly $1\times 10^{-13} {\rm m/s}^2/\sqrt{{\rm Hz}}$ down to 1 microHz. 

	In fact, much of the desired freedom from spurious accelerations needed for LISA has been demonstrated in the laboratory with torsion pendulum measurements.\cite{carbone07}  But, more important, the overall performance of the drag-free system will be demonstrated in the LISA Pathfinder Mission, which is scheduled for launch by ESA in 2014.\cite{antonuccia11,antonuccib11} 

\section{Other scientific benefits from the mission}

	Additional effects such as those arising from non-linear terms in the 00-component of the metric tensor, parameterized by $\beta$, as well as other time delay effects originating in the sun's rotation, can also be measured.  The clock at the L1 point will experience frequency shifts from the earth's potential, solar tidal effects, and second-order Doppler shifts.  Relative to a reference on earth's surface, the fractional frequency shift is about $+6.9\times 10^{-10}$, and is almost all gravitational.  Comparing the clock at L1 with a similar clock on earth's geoid will give accuracies of a few parts per million in a few hours, which is orders of magnitude more accurate than the Vessot-Levine 1976 Gravity Probe A result.  This result is comparable to that expected from the upcoming ACES mission.\cite{much09} 

\section{Postscript}

After the Moriond Meeting we learned about proposals\cite{appourchaux09,braxmaier11} for a mission called ASTROD I, with improved measurement of the gravitational time delay as one of its main objectives. 
In the proposed mission, the time delay measurements would be made
between a drag-free spacecraft in a solar orbit with a semi-major axis of
about 0.6 AU and laser ranging stations on the Earth.  The projected accuracy
for determining the PPN parameter gamma is $3 \times 10^{-8}$.

In these papers the drag-free requirement is given explicitly only over the
frequency range from 0.0001 to 0.1 Hz, and is $3 \times 10^{-14}$ m/s$^2/\sqrt{{\rm Hz}}$ at
0.0001 Hz.  And the only clock frequency stability requirements given are
$1 \times 10^{-14}$ for the clock in the satellite by comparison with ground clocks and
$6 \times 10^{-14}$ stability in the round-trip travel time of roughly 1700 s.  However,
to reach the accuracy goal given for gamma appears to require low levels of
spurious acceleration noise and clock noise down to about 1 microHz or
lower.  Thus it seems possible that quite low spurious acceleration and
clock noise levels at low frequencies actually were implemented in the simulations on which the
ASTROD I accuracy goals are based.

\end{document}